\documentstyle [12pt]{article}
\begin{document}
\begin{titlepage}
\begin{flushright}
ILL-(TH)-92-\#12\\
JUNE 1992\\
CERN-TH.6542/92
\vskip 1.5truecm
\end{flushright}
\begin{center}
{\LARGE Spectroscopy, Scaling and Critical Indices in
Strongly Coupled Quenched QED} \\
\vskip 1.0truecm
{\large Aleksandar Koci\'{c}}\\
\vskip 0.2truecm
{\normalsize  Theory Division, CERN} \\
{\normalsize  CH-1211 Geneva 23, Switzerland} \\
\vskip 1.0truecm
{\large John B. Kogut,  Maria-Paola Lombardo}
\footnote{ on leave from Infn, Sezione di Pisa, Italy} \\
\vskip 0.2truecm
{\normalsize  Department of Physics,}\\
{\normalsize  University of Illinois at Urbana-Champaign,}\\
{\normalsize  1110 West Green Street, Urbana, IL 61801, U.S.A.}\\
\vskip 1.0truecm
{\large K.C. Wang}\\
\vskip 0.2truecm
{\normalsize  School of Physics,}\\
{\normalsize  University of New South Wales}\\
{\normalsize  Kensington, NSW 2203, Australia} \\
\vskip 1.0truecm
\end{center}
\end{titlepage}
\vfill
\newpage
\begin{abstract}
The interplay of spectroscopy, scaling laws and critical indices
is studied in strongly coupled quenched QED. Interpreted as a model
of technicolor having strong interactions at short distances, we predict
the techni-meson mass spectrum in a simplified model of a dynamically
generated top quark mass $M_f$. Our results support the strict inequality
that the techni-sigma mass $M_\sigma$ is less than twice the dynamical
quark mass $M_f$, and confirm that the techni-pion is a Nambu-Goldstone
boson. The level ordering $0 = M_\pi < M_\sigma < 2M_f
< M_\rho < M_{a1} $ is found.An equation of state, and scaling laws are derived
for the techni-meson masses by exploiting correlation length scaling.
The resulting universality relations are confirmed by simulations on
$16^4$, $32\times 16^3$ and $32^4$ lattices. The anomalous dimension
$\eta$ is measured to be approximatively $0.50$ in good agreement with
past lattice simulations and hyperscaling relations, as well as with the
analytic solution
of the quenched, planar gauged Nambu-Jona Lasinio model solved by continuum
Schwinger-Dyson equation techniques.
\end{abstract}
\newpage

\section{Introduction}

{}~~~~Quenched, strongly coupled QED is an interesting, relevant model field
theory~ \cite{REV}, ~\cite{QED}.
Since the photon sector of the theory is free, the model is amenable
to analytic solution and a wealth of interesting dynamics has been found in
its light fermion sector. Chiral symmetry breaking,
a dynamical realization of
the Goldstone mechanism, a rich renormalization group structure, large
anomalous
dimensions, super-critical field phenomena, etc. have been discussed
and illustrated in this context
{}~\cite{REV},~\cite{QED},
{}~\cite{CSB}, ~\cite {MI83_5},
{}~\cite{LLB86_9},~\cite{BLM90}.

Many of the continuum predictions have been
confirmed by lattice simulations which have stressed the theory's
chiral Equation
of State (EOS)~\cite{EOS90}.
 The central purpose of the present paper is to study the theory's
mass spectrum, and relate the spectrum to other fundamental properties of
the theory, such as its critical indices (anomalous dimensions). Since the
model is a simplified version of Technicolor dynamics~\cite{WT}
 and has been advocated
as the dynamical basis of the mass of the top quark, our results should be
of interest to high energy model builders looking beyond the Standard Model
toward the physics underlying the Higgs mechanism~\cite{MI89}.
Using lattice techniques
borrowed from lattice QCD spectrum calculations we shall calculate the masses
of the techni-fermion($f$), pion ($\pi$), rho ($\rho$), sigma ($\sigma$) and
a1 ($a1$) \cite{EXPLO}.
 In the strong coupling phase where chiral symmetry is broken
dynamically, we confirm that the pion is massless and satisfies the constraints
of PCAC. The other states are massive with the ordering
$M_f < M_\sigma < M_\rho < M_{a1}$. In fact, $M_\sigma < 2M_f$ as expected in
a theory with an unscreened attractive electromagnetic force between fermion
and antifermion.

This paper will also pursue the more theoretical goal of relating features of
the theory's
spectrum to universal quantities characterizing its renormalization
group fixed point. The reader should recall that the continuum model must be
parametrized by \underline{two} coupling constants in order for an analytic
renormalization group to exist ~\cite{MI83_5},~\cite{LLB86_9},~\cite{EOS90}.
The electrodynamic coupling $\alpha$  must
be supplemented  by an induced four Fermi coupling $G = G_0 \Lambda^2/\pi^2$
where $G_0$ is the dimensionful four Fermi coupling and $\Lambda$ is an
ultra-violet momentum-space cutoff. Analytical studies of the theory's
Schwinger-Dyson equation reveal that there is a fixed line, in the
renormalization group sense, extending from $\alpha = 0$ to $\alpha_c
= \pi /3$ at $G = (1 + \sqrt{1 - \alpha/\alpha_c})^2$
{}~\cite{LLB86_9}, ~\cite{EOS90}.
The theory
has been particularly well studied at the endpoints of this line.
At $\alpha=0$,
$G = G_c = 4$ the theory reduces to the Nambu Jona-Lasinio (NJL) model
{}~\cite{NJL}.
The cutoff theory breaks chiral symmetry for all $G > G_c$ and when the cutoff
is removed the theory becomes a free field. At all other points along the fixed
line there is an unscreened massless photon which guarantees that the critical
theory is interacting, with a rich mass spectrum. At the other end of the fixed
line ( $G = 1, \alpha = \pi /3$), there is the Miransky-Bardeen-Leung-Love
point where ``collapse of the wavefunction" occurs
\cite{MI89}, \cite{LLB86_9}. Critical indices along
the line of chiral transitions vary continuously as $\alpha$ varies from
zero to $\pi/3$
\begin{eqnarray}
\nu & = &1/ 2 \sqrt{1 - \alpha/\alpha_c} \nonumber \\
\beta_{mag} &  = & (2 - \sqrt{1 - \alpha/\alpha_c}) / 2 \sqrt{1 -
 \alpha/\alpha_c}\\
\delta & = & (2 + \sqrt {1 - \alpha/\alpha_c})/(2 -\sqrt{ 1 - \alpha/\alpha_c})
\nonumber \\
\gamma & = & 1 \nonumber
\end{eqnarray}
with familiar mean field values at $\alpha = 0$ and Miransky's
essential singularities at $\alpha = \alpha_c$
{}~\cite{EOS90}, ~\cite{BLM90}.

A major objective of any numerical lattice approach to this model would be
a verification of the continuum model's phase diagram in the ($\alpha, G$)
plane. Lattice studies which tune the Action and move in a controlled
fashion along the fixed line have not yet been done. Instead, non-compact
quenched QED has been simulated by generating photon field configurations
using a fast Fourier transform and then calculating fermion propagators in that
background using staggered lattice fermion fields ~\cite{QED_REV}.
It has been argued that
the discreteness of the Lattice Action necessarily generates some four Fermi
couplings, so when the lattice fine structure is increased and a chiral
transition is found, one is actually simulating a continuum Action with both
photon exchange and four-Fermi forces. Large lattice, high statistics
studies have measured the critical lattice coupling $\beta = 1/ e^2 = .257(2)$
and critical indices $\delta = 2.2(1)$, $\beta_{mag} = .78(8)$. These values
match the indices $\delta$ and $\beta_{mag}$ in Eq.(1) if ($\alpha, G $) =
($.44\alpha_c,3.06$) \cite{NMF91}.
These rather precise measurements use the framework
of the chiral Equation of State (EOS) as has been discussed in detail before
and will be reviewed below. Briefly, this approach requires that we simulate
the model within its scaling window but not necessarily directly at its
critical point. Simulation data near but not at $\beta_c$ and near but not
at zero fermion mass $m$ should satisfy a chiral EOS and lie on a universal
curve if ``reduced" quantities are plotted. This approach, which is borrowed
from ancient studies of ferromagnetic transitions, has proved to be very
successful.

The model's scaling laws and critical indices can also be found from its
spectrum of states. The development and exploitation of  relations
between spectroscopy, scaling laws and critical indices is a major theme
of this work and new analytic and numerical results will be presented. Our
most successful result will be a scaling law between the chiral condensate
$<\bar\psi\psi>$ and any dynamically generated mass $M_a$,
\begin{equation}
 <\bar\psi\psi> = C_a M_a ^{\beta_{mag}/\nu}
\end{equation}
We shall argue that Eq.(2) should hold anywhere in the theory's scaling region,
i.e. for couplings near the critical point and symmetry breaking fields,
such as the bare fermion mass, which are sufficiently weak. The theoretical
assumptions underlying Eq.(2) consist of (1.) validity of the chiral
Equation of State and (2.) hyperscaling, the idea that the theory's
critical singularities are due to the divergence of a single length scale,
the correlation length. Using our accurate $<\bar\psi\psi>$ data on $16^4$
and $24^4$ lattices and results for $M_\rho$ on $32\times 16^3$ lattices
we shall find that Eq.(2) is well satisfied with the critical indices
\begin{equation}
 \beta_{mag} /\nu = 1.25(1)
\end{equation}
which is in excellent agreement with Eq.(1) which also gives $1.25$ at the
point ($\alpha, G$) = ($.44 \alpha_c, 3.06$) picked out by the lattice
Action. In mean field theory, such as the NJL model, this ratio of critical
indices is unity. Thus, the lattice simulation is correctly accounting
for the effects of an unscreened vector force which renders the theory
non-trivial in the continuum limit.

Other analyses discussed below will give further evidence that the critical
index $\gamma$, which controls the susceptibility divergence, is unity,
and $\nu$, which controls the correlation length, is 0.68(2).

This article is organized into several additional sections. In Sec. 2
we review the relation between the chiral Equation of State, hyperscaling
and critical indices. The approach and results in ref.~\cite{EOS90} play
a crucial role here. In Sec.3 we present our numerical results which are
then used in Sec 4 -- 6 to extract the physics discussed briefly above.
Finally, in Sec.7 we suggest future research.

\section{Scaling and critical indices}
{}~~~~In this section we shall summarize the main theoretical arguments
which proved useful in our analysis of our quenched data. A detailed
exposition can be found in refs.~\cite{EOS90},~\cite{SACHA_NEW}.
The basic philosophy of
refs.~\cite{EOS90},~\cite{SACHA_NEW}
is the use of the chiral Equation of State and hyperscaling,
the assumption that a theory's critical singularities are all due to one
divergent length scale, to obtain scaling laws for low energy quantities
which have an immediate physical significance. This approach will yield
new ways to extract critical indices for low energy properties of the theory,
and will shed interesting insights on these two seemingly different
features of the theory.

We begin with a theory's Equation of State (EOS) written in a standard form
{}~\cite{AMIT}.
Let $t$  denote the deviation from the critical
coupling, and $m$ the symmetry breaking field;
in our application $t = \beta - \beta_c$
where $\beta = 1/ e^2$ ( $e$ = electrodynamic coupling) and $m$ is the
bare fermion mass which explicitly breaks the continuous chiral symmetry of
massless QED. The EOS records the order parameter's response to a change in
coupling and symmetry breaking field,
\begin{equation}
m = <\bar\psi\psi >^\delta f(t/<\bar\psi\psi >^{1/\beta_{mag}})
\end{equation}
Eq. (4) contains the critical indices $\delta$ and $\beta_{mag}$ in standard
statistical mechanics notation. Eq. (4) should hold everywhere in
the theory's scaling region, small $t$ and $m$. It is frequently convenient
to rewrite Eq.(4) in terms of ``reduced" variables and to invert it. Let $y$ be
a reduced symmetry-breaking field and $z$ be a reduced order parameter
\begin{eqnarray}
y & = & m /t^\Delta  \nonumber \\
\\
z & = & < \bar\psi\psi > / t^{\beta_{mag}} \nonumber
\end{eqnarray}
where we have used the conventional definition $\Delta = \delta\beta_{mag}$.
If we divide Eq. (4) by $t^\Delta$ we have
\begin{equation}
y = z ^\delta f ( z ^{-1/\beta_{mag}})
\end{equation}

In other words, the EOS implies that the reduced symmetry-breaking field is
just a function of the reduced order parameter. Upon inverting this dependence,
\begin{equation}
<\bar\psi\psi > = t^\beta_{mag} F(y)
\end{equation}
which is an alternative and familiar form of the EOS. Eqs. (4) and (7)
have been
used extensively to extract the critical indices and
the critical coupling in quenched
lattice QED simulations \cite{NMF91}.
They will be used again with impressive success
in Sec. 4 below.

Our next goal is the determination of ``Equations of State" for other low
energy features of the model, such as its mass spectrum. In order to do that
we assume the correlation length scaling idea, which underlies hyperscaling,
familiar in classical statistical mechanics. We assume that there is a
single macroscopic length scale $\xi$ whose divergence is responsible
for the theory's critical behaviour. $\xi$ sets the scale for low energy
physics (low on the scale of the cutoff, the reciprocal of the lattice
spacing in our case) and one can do dimensional analysis using $\xi$
as a unit of length. In particular, if the field
$\phi$ has scale dimension $d_{\phi}$
and we consider the scaling region of the theory where $\xi$ is much larger
than the lattice spacing, then
\begin{equation}
<\phi> \propto \xi^{-{d_\phi}}
\end{equation}
Futhermore, suppose that the theory has power-behaved critical singularities,
so $\xi$  diverges as $t^{-\nu}$ in the scaling region. Any dynamically
generated mass scale $M_a$ in the model must, by the assumption of correlation
length scaling, behave as $\xi^{-1}$. So, Eqs. (7) and (8) give,
\begin{equation}
<\phi> \propto M_a^{d_\phi} \propto t^{\nu d_\phi} \propto t^\beta_{mag} F(y)
\end{equation}
which implies that $d_\phi = \beta_{mag} / \nu$ and,
\begin{equation}
<\phi> = C_a M_a^{\beta_{mag}/\nu}
\end{equation}
\underline{everywhere}
 in the scaling region. The constants $C_a$ are not controlled
by this scaling argument and are non-universal. We shall find that numerical
data for $<\bar\psi\psi>$ and $M_\rho$ satisfy the scaling law Eq.(10)
beautifully  and yield a relatively precise estimate for $\beta_{mag}/\nu$
which agrees with continuum calculations based on the Schwinger-Dyson
equation \cite{EOS90}, \cite{BLM90}
at the point ($.44 \alpha_c, 3.06$) picked out previously by the lattice
results.

Since the anomalous dimension $\eta$ of the field $\phi$ is defined
as $d_\phi = d/2 - 1 + \eta/2$, Eq.(10) is directly sensitive to
$\eta$ itself. The anomalous dimension $\eta$ vanishes identically
in mean field theory. Therefore, Eq.(10) will prove to be particularly
informative in this and, hopefully, other theories studied in the future.

We shall also find it useful to write Eq.(9) in the form
\begin{equation}
M_a = t^\nu G_a(y)
\end{equation}
and consider ratios of dynamically generated masses,
\begin{equation}
R_{ab} = M_a/M_b = G_{ab}(y)
\end{equation}
which are functions of just one scaling variable $y$. The validity of
this scaling law will also be demonstrated below. In fact, in the following
Sections of this article the three equations Eqs. (7),(10) and (11)
will be tested
and used to extract critical indices. The validity of hyperscaling will be
tested by plotting two mass ratios against one another. The universality
of these plots will provide an independent check of the scaling hypothesis
which was first seen in EOS plots of $<\bar\psi\psi>$. The limiting
($m \to 0, t \to 0$) values of particular mass ratios are universal numbers
and we shall attempt to estimate them below.

In ref ~\cite {SACHA_NEW} the properties of the mass ratios are
discussed in great detail, and several analytic examples are given.
It is observed there that in a wide class of theories in which mesons
are made out of fundamental, pointlike constituents, non-triviality
and compositeness go hand-in-hand. In particular, mesons have non-zero sizes
which guarantee that they interact with one another. Only in the mean field
limit where the size of mesons shrinks to zero and compositeness is lost,
do the theories become trivial. In that case the mass ratio $2M_f/M_\sigma$
is exactly unity. We shall find in quenched QED that $M_\sigma < 2 M_f$ in the
chiral limit, which is indicative of compositeness and non triviality.

We direct the reader to ref.~\cite{SACHA_NEW},~\cite{EOS90} for a more
thorough discussion of the connection between hyperscaling, renormalizability
and non-triviality. We have simply extracted a few portions of a wide class
of relations for our use here and have emphasized only those which shed light
on the non-triviality of the light quark dynamics in quenched lattice QED.

\section{Numerical Results}

{}~~~~We describe here the numerical simulations and the data
analysis. The reader uninterested in these details can skip this
section and go to the subsequent sections devoted to the discussion of
our results.

\subsection{The Simulations}
{}~~~~We have collected  100 , 30 and 248 gauge field configurations on
the $16^4$, $24^4$ and $16^3\times 32$
lattices, respectively , by using the algorithm
introduced in ~\cite{QED4_G}. The algorithm begins in momentum
space and produces the appropriate Gaussian distribution of photons.
Using a Fast Fourier Transform it then generates
a set of dimensionless gauge fields $\theta_\mu$ in coordinate space.
The coupling of the gauge field to the fermion is implemented
by rescaling the gauge fields as
$\theta \to \theta \sqrt\beta$.

This fast algorithm eliminates the correlations between
subsequent configurations (the only source of correlation comes from
the random number generator) and greatly reduces the correlations
between the fluctuations at different spatial sites on the lattice.
Since this algorithm produces independent configurations, our statistical
sample is quite large by present simulation standards.

\subsection{The Chiral Condensate}
{}~~~~The configurations produced on the symmetric lattices have been used
for the computation of $<\bar\psi\psi>$.
We have inverted the staggered Dirac
operator for each value of the fermion mass using a second order
conjugate gradient routine. We used a noisy source for this inversion~
\cite{NOISE}.
On the $16^4$ lattice at each
$\beta$ value (spaced by $\Delta \beta = 0.002$ and ranging from
$\beta = 0.247$ to $\beta = 0.257$) we have accurate
$<\bar\psi\psi>$ values at five fermion masses $m$ equi-spaced from $.001$
to $.005$. We have seen in past studies that accurate data over this
mass range can yield unambiguous, quantitative results for $\beta_c$,
$\delta$ etc. We also recall from past studies that the finite size effects,
as inferred from $24^4$ and $32^4$ simulations, are very small in
$<\bar\psi\psi>$ over this range of $m$. This is further demonstrated by our
new data on the $24^4$ lattice
(presented in Table~\ref{CHI24}), where we explored the same set of masses
$m$ and a very large set of $\beta$'s.

\begin{table}
 \begin {tabular} {||l|l l l l l||} \hline
 \multicolumn{6}{||c||}{\em $<\bar\psi\psi>$     } \\ \hline
 & 0.001 & 0.002 & 0.003 & 0.004 & 0.005\\
  \hline
 0.257&0.0392(9) & 0.0541(7)& 0.0656(7)& 0.0752(7)&0.0833(6)
 \\
 0.255&0.0485(11)&0.0622(8) & 0.0736(8)& 0.0828(7)& 0.0907(7)
 \\
 0.253&0.0583(12)&0.0716(9) & 0.0822(9)& 0.0909(7)&0.0984(7)
 \\
 0.251&0.0694(14)&0.0817(11)&0.0913(10)& 0.0994(8)&0.1065(8)
 \\
 0.249&0.0814(16)&0.0920(13)&0.1006(12)& 0.1081(9)&0.1149(8)
 \\
 \hline
 \end{tabular}

\protect\caption{$<\bar\psi\psi>$ data at various fermion masses and
couplings $\beta$ on the $16^4$ lattice}
\protect\label{CHI16}
\end{table}

\begin{table}
 \begin {tabular} {||l|l l l l l ||} \hline
 \multicolumn{6}{||c||}{$<\bar\psi\psi>$}     \\ \hline
 & 0.001 & 0.002 & 0.003 & 0.004 & 0.005 \\
  \hline
 0.260&0.0296( 14)&0.0436( 12)&0.0546( 11)&0.0639( 10)&0.0721(  9)
 \\
 0.259&0.0324( 15)&0.0466( 13)&0.0577( 12)&0.0670( 11)&0.0751( 10)
 \\
 0.258&0.0352( 16)&0.0498( 14)&0.0609( 12)&0.0702( 11)&0.0783( 10)
 \\
 0.257&0.0383( 18)&0.0532( 15)&0.0643( 13)&0.0735( 12)&0.0816( 10)
 \\
 0.256&0.0419( 20)&0.0569( 16)&0.0678( 14)&0.0770( 12)&0.0850( 11)
 \\
 0.255&0.0460( 23)&0.0608( 17)&0.0715( 14)&0.0806( 12)&0.0885( 11)
 \\
 0.254&0.0508( 25)&0.0648( 18)&0.0753( 15)&0.0843( 13)&0.0922( 11)
 \\
 0.253&0.0559( 28)&0.0690( 19)&0.0793( 15)&0.0881( 13)&0.0959( 12)
 \\
 0.252&0.0611( 30)&0.0732( 20)&0.0834( 16)&0.0921( 14)&0.0997( 12)
 \\
 0.251&0.0662( 31)&0.0776( 20)&0.0876( 16)&0.0961( 14)&0.1037( 12)
 \\
 0.250&0.0710( 31)&0.0821( 20)&0.0919( 17)&0.1003( 14)&0.1077( 13)
 \\
 0.249&0.0759( 31)&0.0869( 21)&0.0964( 17)&0.1046( 15)&0.1118( 13)
 \\
 0.248&0.0812( 31)&0.0918( 21)&0.1010( 17)&0.1090( 15)&0.1161( 14)
 \\
 0.247&0.0869( 31)&0.0970( 22)&0.1057( 18)&0.1135( 16)&0.1204( 14)
 \\
 0.246&0.0930( 31)&0.1022( 23)&0.1106( 19)&0.1180( 16)&0.1247( 14)
 \\
 0.245&0.0990( 31)&0.1075( 24)&0.1155( 20)&0.1227( 17)&0.1292( 15)
 \\
 0.244&0.1048( 33)&0.1128( 25)&0.1204( 21)&0.1274( 18)&0.1337( 16)
 \\
 0.243&0.1104( 34)&0.1182( 26)&0.1254( 21)&0.1322( 18)&0.1383( 16)
 \\
 0.242&0.1159( 35)&0.1236( 27)&0.1305( 22)&0.1370( 19)&0.1430( 17)
 \\
 0.241&0.1215( 35)&0.1291( 27)&0.1358( 23)&0.1420( 20)&0.1478( 18)
 \\
 0.240&0.1279( 35)&0.1347( 29)&0.1410( 24)&0.1470( 21)&0.1527( 18)
 \\
 0.239&0.1344( 35)&0.1406( 30)&0.1464( 25)&0.1521( 22)&0.1575( 19)
 \\
 0.238&0.1411( 37)&0.1466( 32)&0.1519( 27)&0.1573( 23)&0.1625( 20)
 \\
 0.237&0.1478( 40)&0.1526( 34)&0.1574( 28)&0.1625( 24)&0.1675( 21)
 \\
 0.236&0.1542( 43)&0.1586( 36)&0.1630( 29)&0.1678( 25)&0.1725( 22)
 \\
 0.235&0.1602( 47)&0.1645( 38)&0.1686( 31)&0.1731( 26)&0.1776( 22)
 \\
 \hline
 \end{tabular}
\protect\caption{$<\bar\psi\psi>$ data at various fermion masses and
couplings $\beta$ on the $24^4$ lattice}
\protect\label{CHI24}
\end{table}

\subsection{The Analysis of the Spectrum Data}
{}~~~~The spectroscopy data are from the $16^3\times 32$ lattice where
a wall source was used for the inversion of the Dirac operator
after the appropriate gauge fixing.

We worked at five $\beta$ values, ranging from $0.245$ to $0.265$.
For each $\beta$ we have data for five masses, spaced by 0.001,
and ranging from 0.002 to 0.006.
We have thus two sets of data
in the weak coupling region (at $\beta = 0.260$ and $0.265$ )
where we expect to observe clear
signatures of chiral symmetry restoration.

For hadron operators we use the standard local form. The
propagators in the PSeudoscalar, SCalar, VecTor, PseudoVector
channels are then parametrized (in the region of
{\it large } time) as follows\cite{STA_PRO}:
\begin{eqnarray}
G(\tau) & = &
a [ exp(-M\tau) + exp(-M(N_\tau - \tau)]
+  \nonumber \\
 & & (-1)^\tau\tilde a [ exp(-\tilde M\tau) +
exp(-\tilde M(N_\tau - \tau)]
\end{eqnarray}

Turning to the fermion, its parametrization reads:
\begin{equation}
G(\tau) =  a [ exp(-M\tau) - (-1)^\tau exp(-M(N_\tau - \tau)]
\end{equation}

By fitting to the above forms,
we obtained good estimates of the fundamental particles (i.e. the
lower masses in the  direct channel for the PS , VT sectors, and
in the oscillating channel for the SC, PV sectors).
To keep the (possible)
contaminations from excited states under control we vary the extrema of
our fits: typically we fit for $t_1 \le t \le T/2 ~(2 \le t_1 \le 10)$,
the number of degrees of freedom ranging from 6 to 12.
The results level-off for $t_1 \simeq 7-8$ for the pion and the
fermion (for which we performed 1-particle fits), and for $t_1 \simeq 2-3$
for the other particles (2-particles fits).
In the first case (pion and fermion) the errors
on the mass estimates
were obtained from a jack-knife analysis performed by decimating one
configuration at a time (remember that our configurations are independent).
For the other mesons, we quote the weighted average of the results
obtained with different $t_1$.
We also performed effective 2-mass analyses
by exactly solving the equations for masses and amplitudes on
four subsequent time slices; the results are in full agreement with
the ones we quoted, but the errors are larger. In any case
the agreement of the results obtained by local and global fits
is a nice consistency check on our analyses.

To get a good fit the inclusion of secondary particles was
necessary in the VT, SC,and PV sectors, as already said, but their
amplitudes proved to be very small, and often compatible with zero
considering
the statistical errors.
 In many cases we noticed that the inclusion of a
secondary state greatly improved the quality of the fit
(i.e. reducing the $\chi^2$) while leaving completely
unaffected the  results for the fundamental
states. In some cases the results of the fits
suggest
the existence of a very light particle in the secondary channel.
We confirmed
that the numerical significance of very light states was less in the
present data
as compared to past simulations on smaller lattices. We believe
that any evidence for a very light state is unreliable and is,
in fact, indicative of rather large finite size effects.

Figs.1-3 provide an overview of the results of the fits:
we note from them
that the relative
amplitude of the oscillating channel to the non-oscillating one decreases
while going from weak to strong coupling, and from low to high masses.
Figs. 1 shows a sample
of $\pi$ propagators for various sets of parameters, with
the best fits superimposed.  Fig. 2 contains several typical fits
for the vector channel. We plot there the fit obtained with $t = 5$
as a starting point. Note that the fit predicts the data at small t
(i.e. $t < 5$.) with good accuracy.
Some results for the the sigma are shown in Fig.3 :
 its behaviour is satisfactory in the
strong coupling region. In the weak coupling region  the
situation  is somewhat less controlled,
mostly because of large errors in the large t region,
and we feel that
the results for $\sigma$ at weak coupling should be
considered with some extra care.

Summarizing, the  weak coupling,  low mass region turns out to be rather
difficult to treat, while the results for the fundamental particles
in the critical and strongly coupled region are fully satisfactory.
Tables ~\ref{PIOT} --
{}~\ref{FERT} collect our results.

\begin{table}
 \begin {tabular} {||l|l l l l l||} \hline
 \multicolumn{6}{||c||}{\em $\pi$     } \\ \hline
 & 0.265 & 0.260 & 0.255 & 0.250 & 0.245\\
  \hline
 0.002&0.0909( 17)&0.1090( 18)&0.1273( 18)&0.1419( 18)&0.1479( 16)
 \\
 0.003&0.1189( 13)&0.1402( 16)&0.1589( 16)&0.1745( 18)&0.1794( 13)
 \\
 0.004&0.1458( 13)&0.1668( 13)&0.1871( 16)&0.2012( 16)&0.2070( 15)
 \\
 0.005&0.1685( 13)&0.1917( 15)&0.2109( 15)&0.2243( 15)&0.2298( 13)
 \\
 0.006&0.1911( 13)&0.2133( 13)&0.2322( 13)&0.2450( 16)&0.2500( 12)
 \\
 \hline
 \end{tabular}

\protect\caption{Results for the pion mass from one particle fit}
\protect\label{PIOT}
\end{table}

\begin{table}
 \begin {tabular} {||l|l l l l l||} \hline
 \multicolumn{6}{||c||}{\em $\rho$ } \\ \hline
 & 0.265 & 0.260 & 0.255 & 0.250 & 0.245\\
  \hline
 0.002&0.1653( 29)&0.2044( 42)&0.2520( 51)&0.3231( 83)&0.3934(118)
 \\
 0.003&0.1916( 26)&0.2310( 29)&0.2815( 37)&0.3472( 70)&0.4164( 86)
 \\
 0.004&0.2163( 23)&0.2579( 27)&0.3089( 41)&0.3699( 52)&0.4355( 71)
 \\
 0.005&0.2385( 23)&0.2797( 22)&0.3322( 39)&0.3919( 40)&0.4548( 62)
 \\
 0.006&0.2598( 19)&0.3026( 22)&0.3535( 35)&0.4115( 41)&0.4734( 59)
 \\
 \hline
 \end{tabular}
\protect\caption {Rho masses from a two particle fit }
\protect\label{RHOT}
\end{table}

\begin{table}
 \begin {tabular} {||l|l l l l l||} \hline
 \multicolumn{6}{||c||}{\em $\sigma$ } \\ \hline
 & 0.265 & 0.260 & 0.255 & 0.250 & 0.245\\
  \hline
 0.002&0.1905(220)&0.2223(403)&0.2361(259)&0.2922(282)&0.3418(691)
 \\
 0.003&0.2182(132)&0.2338(111)&0.2681(427)&0.3194(281)&0.3690(536)
 \\
 0.004&0.2287( 82)&0.2625(427)&0.2959(200)&0.3480(410)&0.4028(276)
 \\
 0.005&0.2469( 64)&0.2796(111)&0.3207(184)&0.3801(529)&0.4271(361)
 \\
 0.006&0.2673( 80)&0.3032( 82)&0.3447(137)&0.3950(225)&0.4531(170)
 \\
 \hline
 \end{tabular}
\protect \caption {Results for the sigma mass from a two particle fit}
\protect \label {SIGT}
\end{table}

\begin{table}
 \begin {tabular} {||l|l l l l l||} \hline
 \multicolumn{6}{||c||}{\em $a1$ } \\ \hline
 & 0.265 & 0.260 & 0.255 & 0.250 & 0.245\\
  \hline
 0.002&0.1829( 38)&0.2151( 44)&0.2692( 89)&0.3434(425)&0.3716(135)
 \\
 0.003&0.2057( 31)&0.2441( 50)&0.2850( 40)&0.3638(216)&0.4051(103)
 \\
 0.004&0.2271( 26)&0.2601( 43)&0.3088( 38)&0.3767( 70)&0.4307( 83)
 \\
 0.005&0.2463( 26)&0.2837( 34)&0.3315( 37)&0.3950( 74)&0.4544( 82)
 \\
 0.006&0.2682( 44)&0.3043( 30)&0.3528( 38)&0.4127(138)&0.4741(119)
 \\
 \hline
 \end{tabular}
\protect\caption {a1 masses from a two particle fit}
\protect\label{a1T}
\end{table}

\begin{table}
 \begin {tabular} {||l|l l l l l||} \hline
 \multicolumn{6}{||c||}{\em fermion } \\ \hline
 & 0.265 & 0.260 & 0.255 & 0.250 & 0.245\\
  \hline
 0.002&0.0951( 50)&0.1116( 52)&0.1293( 92)&0.1481(151)&0.1782(231)
 \\
 0.003&0.1071( 40)&0.1244( 44)&0.1375( 96)&0.1623(134)&0.1905(214)
 \\
 0.004&0.1180( 37)&0.1343( 61)&0.1497( 89)&0.1742(122)&0.2022(179)
 \\
 0.005&0.1283( 34)&0.1407( 68)&0.1607( 85)&0.1857(116)&0.2136(166)
 \\
 0.006&0.1383( 33)&0.1506( 66)&0.1714( 83)&0.1963(111)&0.2244(157)
 \\
 \hline
 \end{tabular}
\protect\caption{fermion masses, from one particle fit}
\protect\label{FERT}
\end{table}

\subsection{The Chiral Extrapolation}
{}~~~~We conclude this section by briefly describing the extrapolation
procedure which leads to the values
quoted in Table~\ref{EX} and plotted in Fig.4.
These technical points are important
since they underlie the evidence for the Goldstone behavior
of the pion we will discuss in more detail in the next Section.

A crucial ingredient
for the chiral extrapolation is the knowledge of
the functional dependence of the hadron masses on the bare quark
mass. This is not known a priori, and could depend on $\beta$,
possibly being sensitive to the anomalous dimension developed
while approaching the critical point.
In an attempt to use as little prejudice as possible,
we decided
to fit our data to several different functions of $m$. We considered
quadratic and cubic hadron masses as first and second order polynomials
in $m$.
We also tried a Pade' P[1,1] expression $M= (a + bm) /(1 + cm)$.
To critically evaluate the approach to zero of the extrapolated values
as the coupling approaches the critical point
(or its consistency with zero in the case of Pion), we tried fits both
with and without a constant term.

All the fits clearly favour the standard form
$M^2 = M_0^2 + Am + Bm^2$
for all the mesons. $M_0$ is zero for the pion,
and non-zero for the other particles in the strong coupling region.
The extrapolated data are not quite zero in the weak coupling region as well
because, we believe, of finite size effects which push the masses up.
Non-zero
extrapolated values are expected (and found) in the weak coupling region
for the same reason. It is interesting, and gives us more
confidence in the results, that the $M_0$ values obtained with different
parametrizations are quite similar to one other.

\begin{table}
 \begin {tabular} {||l|l l l l ||} \hline
 \multicolumn{5}{||c||}{\em mesons/fermion  $(m = 0)$}    \\ \hline
 & $\rho$ & a1 & $\sigma$ & f \\
  \hline
 0.245&0.3466( 14)&0.2867(101)&0.2623( 58)&0.1513(  6)
 \\
 0.250&0.2664( 24)&0.3207( 26)&0.2190( 39)&0.1150( 25)
 \\
 0.255&0.1721(105)&0.2274( 50)&0.1547(111)&0.1119( 25)
 \\
 0.260&0.1367( 64)&0.1611( 36)&0.1895(125)&0.0785( 58)
 \\
 0.265&0.0965( 84)&0.1337( 28)&0.1694( 15)&0.0671( 20)
 \\
 \hline
 \end{tabular}
\protect\caption{Extrapolated values according to
$M^2 = M_0^2 + a m + b m^2$}
\protect\label{EX}
\end{table}

In Fig.4 we present an overview of the results:
(fermion, $\sigma$ , $\rho$, a1) ($m = 0$)
in the strong coupling region are plotted vs. $\beta$.
In this plot
we discarded the data
for the a1 at the lowest beta value.
It appeared that $M \simeq 0.4$ is an
upper limit to the masses one can meaningfully study at $\beta = 0.245$
because of finite spacing effects. The finite size effects show up clearly
in the results for the fermion mass near $\beta_c$:
 the fermion masses "flatten" near the critical point
to a value
around $0.1$ . Indeed $1/0.1$ is
close to half the time extent of the lattice, so it is unlikely that
we can reliably calculate lower masses.

All the extrapolated masses are well split in the strong coupling region
and approach zero (with the caveats
discussed above) as $\beta \to \beta_c$.

\section{The Equation of State and the Critical Indices $\gamma,\delta$}
{}~~~~Our previous work on chiral symmetry breaking in quenched QED4
has shown that its critical behaviour is not described by
mean field theory \cite{NMF91}.
In fact, our measurements have all been consistent
with the quantitative predictions of the Schwinger-Dyson
equation defined with a momentum cutoff $\Lambda$ and the
hyperscaling relations between critical indices
{}~\cite{EOS90},~\cite{BLM90}.
Our purpose here
is to confirm those results using new, more accurate data and to
extract the critical index $\gamma$, which controls the susceptibility
divergence in the critical region.
A practical
way to find $\gamma$ proceeds through the chiral equation of state,
\begin{equation}
<\bar\psi\psi>/m^{1/\delta}= f(\Delta \beta /<\bar\psi\psi>^ {1/\beta_{mag}})
\end{equation}
where $\delta$ and $\beta_{mag}$ are the usual critical indices discussed
in Sec. 2 above, $\Delta \beta = \beta - \beta_c$ and $f$ is a universal
function.
Our past low mass $<\bar\psi\psi>$ data has given $\beta_c = 0.257(1)$,
$\delta = 2.2(1)$ and $\beta_{mag} = .78(10)$. These results were also
obtained by combining Lanczos data for the spectral density of
$<\bar\psi\psi>$ with conjugate gradient data for its bare fermion
mass dependence. Here we used the conjugate gradient method to calculate
$<\bar\psi\psi>$ and FFT's to generate independent background photon
configurations\cite{QED4_G}.
Our $<\bar\psi\psi>$ data is presented for 100 independent
gauge field configurations on a $16^4$ lattice in Table~\ref{CHI16}.

The data in Table~\ref{CHI16} picks out $\beta_c = 0.257(1)$ very clearly.
Recall that for $\beta = \beta_c$ the order parameter
$<\bar\psi\psi>$ should scale as
\begin{equation}
<\bar\psi\psi> = Am^{1/\delta}    (\beta = \beta_c)
\end{equation}
We test for a simple power law by plotting $\log <\bar\psi\psi>
vs. \log m $ for $\beta = 0.257$ and $0.255$ in Fig.~5. Note
that the $\beta = 0.255$ data indicates that $<\bar\psi\psi>$
at $m = 0.001$ is too large to admit a good power law fit
for $m$ ranging from $0.005$ to $0.001$. However, the power-law fit for
the $\beta = 0.257$ data is excellent and the slope of the line
in Fig.~5 yields $\delta = 2.15(5)$ in agreement with our earlier work.
Inserting these values for $\beta_c$ and $\delta$ into the chiral
equation of state (1) we can ask whether there is a particular
$\beta_{mag}$ which yields a universal curve $f$, i.e. can
all the data of Table~\ref{CHI16} at various $\beta$ and $m$
values be plotted on a single scaling curve? Recalling the
definition
$\gamma = \beta_{mag}(\delta - 1)$  then gives us the susceptibility index
of interest. In Fig.~6 we show the chiral equation of state for $\beta_c
= .257$, $\delta = 2.20$ and $\beta_{mag} = .833$ The success
of scaling hypothesis is very impressive. This value of $\beta_{mag}$
yields $\gamma$ precisely equal to $1.00$.
This result is consistent with, but more accurate than, our past analyses
and is in agreement with the Schwinger-Dyson prediction of
$\gamma = 1.00$. We discussed the theoretical significance of
this result in Sec. 2 above.

The almost perfect scaling seen in Fig.~6 deteriorates rapidly as
$\gamma$ deviates from unity. The dispersion in $f$
exceeds the statistical errors (several representative error bars are shown
in Fig.~6a) if $\gamma$ exceeds $1.05$ or if $\gamma$ falls below $.975$, so
we have determined
\begin{equation}
  \gamma = 1.00 + 0.05 -0.03
\end{equation}

We can exploit the scaling form of EOS also by writing
\begin{equation}
m/<\bar\psi\psi>^{\delta}
= f(
(\beta_c - \beta)/<\bar\psi\psi>^{1/\beta_{mag}})
\end{equation}
On the basis of the very detailed analysis of ref.~\cite{EOS90}
we can predict that the universal function $f$ is straight line.
In fact we know  that the data for the chiral
condensate satisfy
\begin{equation}
m = a(\beta_c - \beta) <\bar\psi\psi> + b < \bar\psi\psi> ^\delta
\end{equation}
which  confirms the theoretical expectation $\gamma = 1$. The previous
equation can be restated like this
\begin{equation}
m /<\bar\psi\psi>^\delta =
a (\beta - \beta_c)<\bar\psi\psi>^{1-\delta} + b
\end{equation}
We can then substitute into the EOS
\begin{equation}
a (\beta - \beta_c)<\bar\psi\psi>^{1-\delta} + b
= f((\beta_c - \beta)/<\bar\psi\psi>^{1/\beta_{mag}})
\end{equation}
So f must be a linear function
with $1/\beta_{mag} = 1 - \delta$, i.e. $\gamma = 1.0$.
It is interesting to check that $f$ actually is a straight line by
looking at the data
on the $24^4$ lattice, shown in Table~\ref{CHI24}, which explores a wide set of
$\beta$'s.  The linear behaviour of Fig.~6b is very clear, and the conclusion
$\gamma = 1.0$ receives further support.

\section{Scaling and the Critical Indices $\eta$ and $\nu$}
{}~~~~As discussed in Sec.2, we can obtain estimates of $\nu$ and $\eta$
by exploiting the relation
\begin{equation}
<\bar\psi\psi> = CM_\rho^{(d/2 - 1 + \eta /2)} = CM_\rho^{(\beta_{mag}/\nu)}
\end{equation}
which should hold everywhere in the scaling region. Moreover, we can test
the consistency of $\nu$ by plotting $M_\rho / t^\nu$ vs
$m/t^{\beta_{mag}\delta}$ as suggested by the equation
of state for the masses
\begin{equation}
M_\rho = t^{\nu} g(m/t^{\beta_{mag}\delta} )
\end{equation}
We will see that the estimates of $\nu$ obtained by using the EOS are
by far more convincing that the one based on the direct fit at $m = 0$
\begin{equation}
M_\rho = A(\beta - \beta_c)^\nu
\end{equation}
which is however consistent with the previous ones.

\subsection {$<\bar\psi\psi>$ vs $M_\rho$ in the Scaling Region}

Consider the scaling law,
\begin{equation}
<\bar\psi\psi> \propto M_\rho^{\beta_{mag}/\nu}
\end{equation}
reviewed in Sec.2 above.
Since mean field theory predicts that $\beta_{mag} = \nu$, deviations
from linear dependence of $<\bar\psi\psi>$ on
$M_\rho$ provide clear evidence of a non-trivial critical point.
The Schwinger-Dyson prediction for the continuum model is
$\beta_{mag}/\nu = 1.25$ if $\delta = 2.2$, as determined from our
Equation of State fit in Sec.3.

We took $<\bar\psi\psi>$ data from the $24^4$ run. They overlap
with the spectroscopy data at $\beta$ = ( $0.260$, $0.255$, $0.250$ ,
$0.245$), $m$ = ($0.002$, $0.003$, $0.004$, $0.005$).
Note that $\beta = 0.260$ lies
in the weak coupling region. However since the transition is rounded, the
data obtained at this $\beta$ value
can be (tentatively) added to the data in the strong coupling
region . We analyzed data sets both with and without the $\beta = 0.260$
points and confirmed that our fits did not change much.

Fig.~7 shows a log-log plot of  $<\bar\psi\psi>$ versus $log~(M_\rho)$,
All the points lie on the same universal plot
(The only small deviation
is at $\beta = 0.260$, at the smallest mass value.) We stress that this
is an highly non-trivial test of Eq. (25),  considering that
the data we used are from
completely independent simulations.
The straight line
is our best fit with all the data included : $<\bar\psi\psi> = M_\rho^{1.25}$.

We fit both $M_\rho$ vs $<\bar\psi\psi>$  and
$<\bar\psi\psi>$ vs $M_{\rho}$,
with and without the points at $\beta = 0.260$.
The four fits give, respectively
$\beta_{mag}/\nu = 1.265, 1.275, 1.243, 1.249 $.
Assuming that $\beta_{mag} = 0.86(3)$
we get for $\nu$:
$ \nu = 0.680(25)$,
$0.674(25)$,
$0.691(25)$,
$0.688(25)$.
These values are different from the mean field prediction,~~$\nu = 1/2$, but
it is not easy  to estimate the errors associated
with this fitting procedure. Rather than attempting this, we prefer to show the
 inconsistency of our results with the mean field
prediction in a more direct fashion. In Fig.~8
we plot $<\bar\psi\psi>^x$ vs $M_\rho$ for
different $x$ values. The solid line corresponds to our best fit, the dashed
ones to the mean field result $x = 1$, the dotted line to $\beta_{mag} = 0.86$
 and
$\nu = 0.5$. The sensitivity of the quality of the fits to the choice of the
exponent $x$ is clear, and thus the incompatibility of our results with the
mean
field prediction is established.

We see that the anomalous  dimension $\eta$ which follows is $ \simeq 0.5$.
This is a remarkably large value. Nonetheless, it is consistent with
the hyperscaling relation between $\delta$ and $\eta$:
\begin{equation}
(6 - \eta)/ (2 + \eta) = \delta
\end{equation}
with $\delta = 2.2$ as determined in Sec.3 above.

In conclusion, our data rule out mean field behaviour.
They predict $\nu = 0.68(3)$ if $\beta_{mag} = 0.86(3)$.

\subsection{The Equation of State for the Masses and Correlation Length
Scaling}
As stated in Sec.2, we can check our estimate of $\nu$ by using the data
of the masses alone. This is done in Fig.~9 where we plot
$M_\rho/(\beta_c - \beta)^\nu$ vs
$m/(\beta_c - \beta)^{\beta_{mag}\delta}$,
for $\nu = 0.67$, $\beta_{mag} = 0.833$, $\delta = 2.20$. We use only points
at strong coupling.
The scaling hypothesis works very well.
Again, we checked that the scaling behavior
deteriorates if we select mean field exponents.
The equation of state for the masses can be studied for $m = 0$. In this
case we get $M_\rho(m=0) = g(0)t^\nu$ and
a direct measure of $\nu$ can be obtained by fitting
$M_\rho(m = 0)$ as a power of $\beta - \beta_c$.
Unfortunately
the quality of our data does not allow a careful estimate of $\nu$
in this way because of uncertainties in the extrapolation to $m = 0$.
The point at $\beta = 0.255$ can be affected by finite volume effects,
and the two remaining points alone are not sufficient to allow a
meaningful, controlled estimate of $\nu$. It is necessary to
work very close to the transition, because previous studies have shown
that in the strong coupling region mean field theory functional
dependences are expected. With these caveats it is still interesting
to get an independent estimate of $\nu$ even if it is crude.
Our best fit (which is shown in Fig.~10) gives
$M_\rho (m = 0)$ = $A(\beta - 0.260)^{0.675}$ which is nicely consistent
with the more sophisticated analysis done in Sec.5.1 above.
It also interesting to note that the ratio of the extrapolated masses for
the $\rho$ and the fermion is constant within statistical errors:
we obtained
$  M_f  / M_\rho  =  0.434 (5)$ at $\beta = 0.245$, and
$  M_f / M_\rho =  0.426 (2)$,
at $\beta = 0.250$,
in reasonable mutual agreement.
This agreement again supports scaling behaviour.

\subsection{Critical indices,  summary}
{}~~~~We present in Table~\ref{SUM} an overview of the critical exponents and
the hyperscaling relations. Let us recall that the estimates for $\beta_{mag}$,
$\delta$, $\eta$, $\nu$ come directly from independent numerical analysis,
and $\gamma$ has been computed according to $\gamma = \beta_{mag} (\delta -
1)$.
So, in the first column we quote the  results of the numerical analysis,
and in the second column the results obtained from the hyperscaling
relations assuming
the other exponents as input. (When we do not have a safe estimate
of the errors we simply quote the central values.) The third column
shows the mean field results (we recall that $\gamma = 1.0$ is also
the theoretical prediction).
We summarize in the last four rows of Table~\ref{SUM}
 the hyperscaling relations we used.

\begin{table}
 \begin {tabular} {||l|l l l||} \hline
 \multicolumn{4}{||c||}{\em Critical indices } \\ \hline
 Index & Result from the simulation & Result from HS  & MF\\
  \hline
 $\beta_{mag}$  &0.86(3)&0.86(6) &0.5\\
 $\gamma = \beta_{mag} (\delta - 1)$ &1.00(5)&   &1.0\\
 $\delta$ &2.2(1)&2.16&3.0\\
 $\eta$   &0.5(1)&0.5&0.0\\
 $\nu$    &0.675&0.68(3)&0.5\\
 $ - 4\nu + 2\beta_{mag} + \gamma$             &0.1(1)  &0.0&0.0\\
 $(2-\eta)\nu/\gamma$                    &1.1(1)  &1.0&1.0\\
 $(2\nu - \gamma)/(\nu\eta  )$           &1.1(1)  &1.0&1.0\\
 $(6 - \eta)/(2 + \eta) - \delta$&0.13(20)&0.0&0.0\\
\hline
\end{tabular}
\protect\caption{Critical indices and relations among them:
in the first column results
from the simulation, in the second column
the HS(hyperscaling) prediction assuming the
other indices as input for the single index entries, in the third column
the mean field prediction (d=4)}
\protect\label{SUM}
\end{table}

We believe that these results support a picture of non-trivial critical
behaviour consistent with hyperscaling,  with a large anomalous
dimension $\eta$.

\section{Spectroscopy: Results}
{}~~~~We devote this Section to the discussion of
the Spectroscopy results presented in
Tables. ~\ref{PIOT} --
{}~\ref{FERT}.

\subsection{The Goldstone Character of the Pion}

{}~~~~Our first task is to check that the pion of quenched QED has all the
attributes of a Nambu-Goldstone particle associated with spontaneous
breakdown of a continuous chiral symmetry. These properties include 1.
the pion should be massless in the strong coupling phase where chiral
symmetry is broken, 2. the square of the pion mass should be proportional
to the bare mass of the fermion if that mass is sufficiently small, and 3.
PCAC relations should hold for $f_\pi$,$m$,$M_\pi$, and
$<\bar\psi\psi>$.

Using the data in Table 3 we checked that $M_\pi ^2 =
Am + B m^2$  works well (see Section. 2 for more details) :
this is the usual parametrization for the
dependence of the pion mass on the bare quark mass assumed in QCD
when the quark masses are not really {\it small}.  We checked that if
we added a constant term to the fit, its best value
was consistent with zero.

We also looked at the dependence of the pion mass on the quark mass  in
a slightly different way, i.e by fitting $M_\pi = A m ^x$ (Fig.~11).
We see that in the strong coupling region the slopes of the
log-log plots for different $\beta$'s are
quite similar, and roughly consistent with a square root dependence
of $M_\pi$ on $m$ (the actual values of the slopes are 0.478, 0.495,
0.547 0.610 0.678 from $\beta = 0.245$ to $\beta = 0.265$).
In conclusion, we observe  only slight deviations
from $M_\pi \simeq \sqrt(m)$ in the strong coupling region.
These deviations can be eliminated by letting
the exponent vary slightly, or,
more conventionally, by adding a second order
term in the $m$ expansion. In the weak coupling region the deviations are
somewhat more important, but they can be controlled in the same way.
In conclusion, the pion has the properties of an ordinary Goldstone
boson.

To complete the discussion on PCAC, we show in Table~\ref{FPI}
(and in Fig.~12) the
results for
$$f^2_\pi = 2m <\bar\psi\psi> / m^2_\pi$$

\begin{table}
 \begin {tabular} {||l|l l l l ||} \hline
 \multicolumn{5}{||c||}{$f^2_\pi$}     \\ \hline
 & 0.002 & 0.003 & 0.004 & 0.005 \\
  \hline
 0.260&0.0147(  6)&0.0167(  5)&0.0184(  4)&0.0196(  4)
 \\
 0.255&0.0150(  6)&0.0170(  4)&0.0184(  4)&0.0199(  3)
 \\
 0.250&0.0163(  5)&0.0181(  5)&0.0198(  4)&0.0214(  3)
 \\
 0.245&0.0197(  6)&0.0215(  5)&0.0229(  4)&0.0245(  4)
 \\
 \hline
 \end{tabular}
\protect\caption{$f_\pi$ as a function of $\beta$ and $m$}
\protect\label{FPI}
\end{table}
The data for
$<\bar \psi \psi>$  overlap
with the spectrum data for four mass values and four $\beta$ values.
There is sufficient data to see that
$f^2_\pi $ is nicely linear in $m$ (the lines in the plot are to
guide the eye), and, most importantly, extrapolates to a non-zero value.

\subsection{Mass Ratios and Level Ordering}

{}~~~~We discuss now the behaviour of some relevant mass ratios: we present
them in a scale invariant form in Figs.~13-16 , where we plot only the
points belonging to strong coupling region: the squares are for
$\beta = 0.255$, the diamonds for $\beta = 0.250$, the crosses for
$0.245$ .
Note that since we used the same values
of quark masses
at the different $\beta$ values the physical mass region we explore
changes with $\beta$: near the critical point the physical masses
are bigger.

Let us first discuss the behaviour of $\rho / fermion$. The plot (Fig.~13)
shows
nice scaling (i.e. all the points fall on top of each other). Some deviations
are observed near the critical point
(but still the data points are on the same line within
errors), which would suggest a $\rho$ lighter than twice the
fermion mass in the chiral limit. However, practical difficulties,
discussed in Sec.3 above, which effect
the fits of the fermion propagator at small mass near criticality,
together with the good agreement of the scaled data at $\beta = 0.245$
and $\beta = 0.250$ , suggest that  the leftmost point at $\beta = 0.255$
be discarded so that $M_\rho (m = 0) \ge 2 M_f(m = 0)$.

The data for $\sigma/ fermion$ (Fig.~14) are particularly interesting
for technicolor applications \cite{WT} and show nice scaling as well
(unfortunately the statistical errors are large). The results
in the figure suggest that in the chiral limit the sigma mass
is less than twice the dynamical mass of the fermion.
However, the deviations from the NJL result (that the sigma is twice as
heavy as the fermion) are not
very large and are partially masked by
statistical errors. As discussed in Sec.2 above, and in~\cite {SACHA_NEW}
a deviation from the
NJL result for this ratio implies non-trivial behaviour.

The $a1$ and $\rho$ (Fig.~15) are almost degenerate
in the large mass region. Some deviations
to scaling are observed at $\beta = 0.245$: they can be ascribed both to
true scaling violations, and
finite spacing effects (note that the magnitude of finite
spacing effects depend on masses in lattice, not in physical units).
The figure suggests that the degeneracy of the a1 and the $\rho$ is
resolved in the chiral limit, and the a1 is the heavier state.

The last plot shows the ratio between the $\sigma$ and $\rho$ (Fig.~16),
and no deviations in scaling are observed.

In conclusion, our data show some evidence for asymptotic scaling
( continuum behaviour), and suggest the level ordering
$0 \le \sigma \le 2 \times fermion \le \rho \le a1$.

\section{Conclusion}
{}~~~~The results presented here on spectroscopy and critical indices of
quenched lattice QED are in good agreement with analytic calculations
based on the continuum formulation of the model. The hypothesis of
correlation length scaling seems well verified by our data which
shows universal behaviour wherever it is expected. Our analysis of
the relation $<\bar\psi\psi> \propto M_\rho^{\beta_{mag}/\nu}$ was
particularly precise and gave a value $\beta_{mag} / \nu = 1.25(1)$
in excellent agreement with our past interpretation of lattice studies
of the model's EOS -- that the lattice Action corresponds to the
point $(\alpha, G) = (.44 \alpha_c , 3.06)$ on the renormalization
group fixed line of the continuum model.

Our spectroscopy results illustrate the type of techni-meson masses
one should expect in a class of models of the top quark. It was particularly
interesting that our data favored the inequality $M_\sigma < 2M_f$ expected
of a theory with unscreened vector forces rather that the familiar NJL result
that the techni-sigma lies at the two fermion threshold. Earlier analytical
work on quenched planar QED had suggested that the critical point would be
characterized by a massless dilaton. This picture is not supported by our
data. Later analytic work recognized that dilaton current conservation
is broken by hard operators and the techni-sigma has no reason to be
light. ``Realistic" Technicolor and Extended Technicolor models of the
dynamically generated mass of the top quark are not left-right symmetric
like the lattice model considered here. There are proposals in the literature
for incorporating chiral fermions into lattice regulated theories. If any
of these proposals prove theoretically sound, and computationally
practical, then it would be interesting to generalize the calculations
of this paper to other, more realistic models of the dynamics behind the
Standard Model.

Instead of reiterating the results of Secs. 4-6 above, we close with a short
discussion of the puzzling features of the quenched model and our simulations.
Since the underlying photon dynamics is perfectly free, the quenched model
does not satisfy all the consistency conditions of a bona-fide field theory.
However, we do not yet clearly understand the model's limitations. In fact,
we found that its light fermion-meson sector satisfies all the scaling laws
expected of complete theories. One peculiarity however is the observation
that the simulations in the weak coupling phase do not display all the
expected results. For example, since chiral symmetry is not broken at
weak coupling the pion and the sigma should be degenerate, and the $\rho$ and
$a1$ also should be degenerate here. No evidence was found for this symmetry
restoration in our spectroscopy data. This failure could just be technical
-- we observed that the usual fitting procedures were not compelling at
weak coupling, so other functional forms of the meson propagator's
spatial dependence should be investigated. Another problem with the data
concerns the dynamical fermion mass. We found that this mass ``flattens"
out as the bare fermion mass is taken small, so it does not have the same
systematic behaviour as the meson masses. We suspect that this result
is due to the fact that the fermion carries an unscreened charge and therefore
experiences the long range interactions of the free photons. Under these
conditions its propagator probably suffers from especially large finite
size effects which push the dynamical mass estimates up when the bare fermion
mass becomes too small. It would be useful to pursue this point and obtain
better control over the fermion mass since it plays such a central role
in theoretical developments.

And finally, lattice investigations of quenched QED must be generalized
to move away from the critical point $(\alpha, G) = (.44\alpha_c,3.06)$.
We must learn how to tune the lattice Action to move along the critical
line predicted by the analytic calculations. If this cannot be done, then
there may be some ingredient missing in the analytic calculations, or
some unsuspected limitation in the lattice approach. From a physics
perspective,
it would be particularly interesting to tune the lattice Action and watch
the critical indices vary continuously as we move from the free field
NJL point, to the super-critical, essential singularities of the Miransky
point.

\vskip 3 truecm
{\bf Acknowledgement}

This work was supported in part by the National Science Foundation,
NSF-PHY 92-00148.

\newpage

{\bf Figure
Captions}
\begin{description}

\item[1]
a)Pion propagators at $\beta = 0.260$ as a function of euclidean
time for the five mass values. The solid lines superimposed are
the results of the fits  for $5 \le t \le 16$. Fig.~1b is the same as
Fig.~1a, at $\beta = 0.250$.

\item[2] a), b)  As Figs.~ 1, for the $\rho$.

\item[3]
a)$\sigma$ propagators at $\beta = 0.250$ as a function of euclidean
time for $m = 0.002$. The solid line superimposed are
the result of the fits  for $5 \le t \le 16$.
Fig.~3b is the same as Fig.~3a, for $m = 0.006$.

\item[4]
Results for the chiral limit for $ ( fermion, \sigma, \rho, a1 ) $
( circles, diamonds, squares, crosses )
(obtained according to
$M^2 = M_0^2 + am + b m^2$)   as a function of $\beta$.

\item[5]  $\log <\bar\psi\psi>$ vs $\log m$ at $\beta$ values
$0.257$ (squares) and $\beta = 0.255$ (circles). Data from
table 1.

\item[6]
a) Chiral equation of state for $\beta_c = 0.257$, $\delta = 2.2$
and $\beta_{mag} = 0.833$  on the $16^4$ lattice.
$(\beta_c - \beta)/
<\bar\psi\psi>^{1/\beta_{mag}}$
is plotted versus $<\bar\psi\psi> /
m^{1/\delta}$ . We plot only points in the strong coupling region.
b) Chiral equation of state on the $24^4$ lattice for the same $\beta_c$,
$\delta$ and $\beta_{mag}$.$m/<\bar\psi\psi>^{\delta}$ is plotted versus
$(\beta_c - \beta)/<\bar\psi\psi>^{1/\beta_{mag}}$.
All the points are shown.

\item[7]
 Log-log plot of $<\bar\psi\psi>$ vs $M_\rho$. (Crosses, diamonds, circles,
squares) are for $\beta = (0.260, 0.255, 0.250, 0.245)$.

\item[8]
$ <\bar\psi\psi>^{\beta_{mag} /\nu} vs M_\rho$.
$\beta_{mag}/\nu = (1, 0.86/0.67,
0.86/0.5)$ (dash, solid, dot line)

\item[9]
$M_\rho /(\beta_c - \beta)^\nu$ vs. $m/(\beta_c - \beta)^{\beta\delta}$
for $\nu = 0.67$, $\beta = 0.84$, $\delta = 0.22$.

\item[10]
$M_\rho (m = 0)$ vs. $\beta$. The solid line is
the best fit $M_\rho = A(\beta - 0.260)^{0.675}$

\item[11] $\log M_\pi$ vs. $\log m$ for $\beta = (0.245, 0.250, 0.255,
0.260, 0.265)$ (top to bottom). The straight lines superimposed are fits to the
relation $M_\pi = A m^x$

\item[12] $f^2_\pi$ as a function of $m$  for $\beta = (0.245, 0.250, 0.255,
0.260)$.

\item[13] Scale invariant plot for $(\rho / fermion)^2$ vs $(\pi/\rho)^2$.
(Squares, diamonds, crosses) are for $\beta = (0.255, 0.250, 0.245)$.

\item[14] Scale invariant plot for $(\sigma / fermion)^2$ vs $(\pi/\rho)^2$.
(Squares, diamonds, crosses) are for $\beta = (0.255, 0.250, 0.245)$.

\item[15] Scale invariant plot for $(a1 / \rho)^2$ vs $(\pi/\rho)^2$.
(Squares, diamonds, crosses) are for $\beta = (0.255, 0.250, 0.245)$.

\item[16] Scale invariant plot for $(\sigma / \rho)^2$ vs $(\pi/\rho)^2$.
(Squares, diamonds, crosses) are for $\beta = (0.255, 0.250, 0.245)$.

\end{description}

\end{document}